\documentclass[pra,aps,showpacs,twocolumn,superscriptaddress,nobibnotes]{revtex4-1}
\usepackage{amsmath}
\usepackage{amssymb}
\usepackage{graphicx}
\usepackage{epstopdf}
\usepackage{mathtools}
\newcommand{\ket}[1]{|#1\rangle}
\newcommand{\bra}[1]{\langle #1|}
\usepackage[bookmarks=false]{hyperref}
\hypersetup{colorlinks=true,citecolor=blue,linkcolor=blue,urlcolor=blue,pdfstartview=FitH,bookmarksopen=true}
\usepackage{xcolor}
\usepackage{soul}
\usepackage{times,txfonts}
\setstcolor{blue}
\setulcolor{red}

\begin{document}

\title {Higher-order protection of quantum gates: Hamiltonian engineering \\
coordinated with dynamical decoupling}
\author{P. Z. Zhao}
\affiliation{Center for Quantum Technologies, National University of Singapore, Singapore 117543, Singapore}
\author{Tianqi Chen}
\affiliation{Center for Quantum Technologies, National University of Singapore, Singapore 117543, Singapore}
\affiliation{Department of Physics, National University of Singapore, Singapore 117551, Singapore}
\author{Sirui Liu}
\affiliation{Department of Physics, National University of Singapore, Singapore 117551, Singapore}
\affiliation{Department of Chemistry, National University of Singapore, Singapore 117543, Singapore}
\affiliation{Joint School of National University of Singapore and Tianjin
University, International Campus of Tianjin University, Binhai New City,
Fuzhou 350207, China}
\author{Jiangbin Gong}
\email{phygj@nus.edu.sg}
\affiliation{Center for Quantum Technologies, National University of Singapore, Singapore 117543, Singapore}
\affiliation{Department of Physics, National University of Singapore, Singapore 117551, Singapore}
\affiliation{Joint School of National University of Singapore and Tianjin
University, International Campus of Tianjin University, Binhai New City,
Fuzhou 350207, China}

\date{\today}

\begin{abstract}
Dynamical decoupling represents an active approach towards the protection of quantum memories and quantum gates.  Because dynamical decoupling operations can interfere with a system's own time evolution, the protection of quantum gates is more challenging than that of quantum states. In this work, we put forward a simple but general approach towards the realization of higher-order protection of quantum gates and further execute the first cloud-based demonstration of dynamical-decoupling-protected quantum gates at the first order and the second order. The central idea of our approach is to engineer (hence regain the control of) the gate Hamiltonian in coordination with higher-order dynamical decoupling sequences originally proposed for the protection of quantum memories. The physical demonstration on an IBM quantum processor indicates the effectiveness and potential of our approach on noisy intermediate scale quantum computers.
\end{abstract}

\maketitle

\section{Introduction}

A crucial challenge in implementing quantum computation is to overcome the decoherence induced by both internal and external environments of a quantum computer.
Quantum error correction protocols make it possible to reliably store and process quantum information over long time scales, conditional upon the fact that the noise level in individual quantum gates is below a certain threshold \cite{Knill,Preskill}. However, reaching the required low-noise threshold for fault-tolerant quantum computation is still challenging for physical implementations today.  It is thus highly beneficial to execute active protection schemes for individual physical qubits to enhance quantum gate fidelity.

As an active approach, dynamical decoupling (DD) aims to average out the system-environmental interaction and hence protects either quantum memories or quantum gates \cite{Viola,Viola1999}. For quantum memories, periodic DD (PDD) \cite{Viola2003} eliminates the effect of the system-environment interaction to the first order (e.g., in the sense of a Magnus expansion), whereas concatenated DD (CDD) \cite{Khodjasteh} and Uhrig DD (UDD) \cite{Uhrig,Yang,West2010,Wang2011} can offer higher-order elimination of decoherence \cite{Gong,Gong2010,Liu2010,Liu2011}. Experimental advances on DD-based quantum memory protection have been highly successful \cite{Biercuk,Du,Ryan,deLange,Wang,Alvarez}, with the recent benchmarking studies of DD carried out on actual noisy quantum computers \cite{Tripathi,Ezzell}.  However, most of this experimental progress on quantum memory protection has not been translated to the protection of quantum gates, because DD operations often interfere with quantum gate operations, e.g., DD pulses tend to freeze the system's own time evolution.

One solution towards the DD protection of quantum gates started with choosing a dagger-closed DD group algebra, under which the system Hilbert space will include a subsystem where DD pulses applied to the system act like an identity operator on this subsystem \cite{Zanardi,Viola2000}. Then, this subsystem can be used to encode a qubit and the commutant of the group algebra can be used to perform computation under DD. However, the implementation of the coupling between such subsystem-encoded qubits is rather complicated, making this route an experimental challenge.
The second solution is to choose certain driving Hamiltonian commutable with DD operations \cite{Byrd,Lidar2008,West}, the so-called ``non-interference" condition.
However, it is difficult to satisfy the non-interference condition when performing universal quantum gates on general single physical qubits.
It is also highly nontrivial to construct the commutable driving Hamiltonian when combining DD with other noise-mitigating strategies such as quantum error correction \cite{Ng} and decoherence-free subspaces \cite{Zhao}.
Therefore, this stimulating route needs extra physical-qubit resources to encode logical qubits, the implementation of which is often based on particular dynamical symmetries or specific qubit-qubit interactions to meet the non-interference requirement \cite{Xu,Wu2020,Wu}.
The third solution is based on the average Hamiltonian theory, where the gate operations are inserted between equidistant DD pulses such that $U=U_{N+1}\prod^{N}_{n=1}P_{n}U_{n}$. Here, $P_{n}$ denotes the propagator of DD pulses and $U_{n}=\exp(-iH_{C_{n}}\tau_{n})$ represents the time evolution between the DD pulses, induced by the gate Hamiltonian $H_{C_{n}}$.
The explicit form of the driving Hamiltonian $H_{C_{n}}$, which yields the unitary operator $U_{n}$, is determined by the equation $\tilde{H}_{C_{n}}=T^{-1}_{n}H_{C_{n}}T_{n}$ in the so-called toggling frame of the DD sequence, where $T_{n}=\prod^{n-1}_{k=1}P_{k}$ with $T_{1}=T_{N+1}=I$.
This approach typically extends the total gate time and can only realize first-order protection \cite{Suter,Zhang,Souza}.

To fill in the gap between quantum memory protection and quantum gate protection, the dynamically corrected gate (DCG) was proposed as a sophiticated composite quantum gate constructed from primitive building blocks without encoding \cite{Viola2009,ViolaPRA}. DCG may offer higher-order protection with a concatenate design \cite{Viola2010}, but the total number of the required gate operations will be huge and thus there is essentially no experimental progress so far.  By contrast, this work aims to show that higher-order protection of quantum gates can be directly achieved on general physical qubits, based on an experimentally friendly Hamiltonian engineering approach plus well-known protocols designed for quantum memory protection.  Indeed, the central idea of our higher-order protection approach lies in the engineering of the time dependence of the system's driving Hamiltonian \cite{Viola2003}, in coordination with known higher-order DD sequences, such as CDD, UDD, and even nested UDD. As seen below, our engineering of the system's driving Hamiltonian is to act against the unwanted impact of DD pulses on the system. This being done, only the system-environment interaction part is suppressed to higher orders and we recover our control over the system's driving Hamiltonian. The DD protection of quantum memories and quantum gates can thus be unified and applied on an equal footing.

To date one-qubit gates can be executed at very high fidelities \cite{Hart,Barends,Evered}. Our higher-order approach allows to execute one-qubit gates over a much longer duration. The extended execution time for one-qubit gates can be particularly crucial for hybrid quantum systems involving different types of qubits (e.g., the electron and nuclear spins of NV centers in diamond) where different qubits with markedly different operational time scales need to be synchronized \cite{Zhang,Zhang2015,Sar,LiuNC2013}. For two-qubit gates, our approach offers higher gate fidelity using single-qubit gates as the DD pulses.
This indicates a new philosophy towards the design of basic building blocks for two-qubit gates: high-fidelity  two-qubit gates may be composed by a series of available high-fidelity single-qubit gates together with engineered two-qubit gate Hamiltonians.  In addition to numerical experiments to illustrate the performance of our higher-order protection schemes, we perform the first cloud-based demonstration of  DD-protected quantum gates at the first or the second order, verifying the effectiveness of our approach on an actual IBM quantum processor~\cite{IBMQuantum}.

\section{Approach}

Consider a quantum system coupled to its environment with the total Hamiltonian
$H(t)=H_{\rm S}(t)+H_{\rm E}+H_{\rm I}$. Here $H_{\rm S}(t)$ is our system Hamiltonian, $H_{\rm E}$ is the environment Hamiltonian and $H_{\rm I}$ is the system-environment interaction Hamiltonian. Using the one-qubit system as an example, the general interaction between a qubit and
its environment is in the form
$H_{\rm I}=\sigma_{x}\otimes{E_{x}}+\sigma_{y}\otimes{E_{y}}
+\sigma_{z}\otimes{E_{z}}$, where ${E}_{x(y,z)}$ represent the associated environment operators.
To illustrate the underlying principle of our idea, we start with the protection of quantum gates under the PDD. For our purpose, we nest a periodic DD sequence $\{\sigma_{0}\equiv{I}, \sigma_{1}\equiv{\sigma_{x}}, \sigma_{2}\equiv{\sigma_{y}}, \sigma_{3}\equiv{\sigma_{z}}\}$ ($I$ corresponding to no pulse applied) in between the time evolution governed by the driving Hamiltonian.
The total evolution operator is then given by
\begin{align}\label{eq1}
\mathcal{U}
=&\prod^{3}_{k=0}\sigma_{k}
\left[\mathcal{T}e^{-i\int^{(k+1)\tau}_{k\tau}H(t)dt}\right]
\sigma_{k}
\notag\\
=&\prod^{3}_{k=0}
\mathcal{T}e^{-i\int^{(k+1)\tau}_{k\tau}
\sigma_{k}H_{\rm S}(t)\sigma_{k}dt}
e^{-i\left[\sum^{3}_{l=0}\sigma_{l}H_{\rm I}\sigma_{l}+4H_{\rm E}\right]\tau}+\mathcal{O}(\tau^{2}),
\end{align}
where the term with the smallest $k$ is placed at the right most and $\mathcal{T}$ denotes the time ordering. Note that $\sum^{3}_{l=0}\sigma_{l}H_{\rm I}\sigma_{l}=0$, so we have
\begin{align}\label{eq2}
\mathcal{U}=&\prod^{3}_{k=0}\mathcal{T}e^{-i\int^{(k+1)\tau}_{k\tau}\sigma_{k}H_{\rm S}(t)\sigma_{k}dt}{\otimes}e^{-i4H_{\rm E}\tau}+\mathcal{O}(\tau^{2}).
\end{align}
Clearly then, the first-order effect arising from $H_{\rm I}$ is eliminated. It is also seen that the time evolution of the quantum system is strongly influenced by the DD pulses.  Indeed, during each time interval of duration $\tau$, the effective Hamiltonian [the exponentials in Eq.~(\ref{eq2})] is updated by the DD pulses. Let us now unravel a key observation behind our Hamiltonian engineering approach. Denote the effective Hamiltonians altered by DD pulses associated with the time interval $[k\tau,(k+1)\tau]$ as
$H_{k}(t)\equiv\sigma_{k}H_{\rm S}(t)\sigma_{k}$.
As an example, we take the driving Hamiltonian of the quantum system as
$H_{\rm{S}}(t)=\Omega_{x}(t)\sigma_{x}+\Omega_{y}(t)\sigma_{y}$, where $\Omega_{x}(t)$ and $\Omega_{y}(t)$ are time-dependent parameters. One can then obtain
$H_{0}(t)=\Omega_{x}(t)\sigma_{x}+\Omega_{y}(t)\sigma_{y}$,
$H_{1}(t)=\Omega_{x}(t)\sigma_{x}-\Omega_{y}(t)\sigma_{y}$,
$H_{2}(t)=-\Omega_{x}(t)\sigma_{x}+\Omega_{y}(t)\sigma_{y}$,
and
$H_{3}(t)=-\Omega_{x}(t)\sigma_{x}-\Omega_{y}(t)\sigma_{y}$, namely, the DD pulses affect the effective Hamiltonians associated with different time intervals through altering the signs of the parameters $\Omega_{x}(t)$ and $\Omega_{y}(t)$.
To fight against these changes to the system Hamiltonian imposed by the DD pulses, we adjust the parameters for different time intervals as
$\Omega_{x}(t)=\Omega(t)\cos\varphi$ and
$\Omega_{y}(t)=\Omega(t)\sin\varphi$ for $t\in[0,\tau]$,
$\Omega_{x}(t)=\Omega(t)\cos\varphi$ and
$\Omega_{y}(t)=-\Omega(t)\sin\varphi$ for $t\in(\tau,2\tau]$,
$\Omega_{x}(t)=-\Omega(t)\cos\varphi$ and
$\Omega_{y}(t)=\Omega(t)\sin\varphi$ for $t\in(2\tau,3\tau]$,
and $\Omega_{x}(t)=-\Omega(t)\cos\varphi$ and
$\Omega_{y}(t)=-\Omega(t)\cos\varphi$ for $t\in(3\tau,4\tau]$.
Because $\Omega(t)$ is identified as the Rabi frequency of the qubit Hamiltonian and $\varphi$ represents a phase of the driving field coupling the two levels of the qubit, the above adjustment can be implemented by only quenching the phase parameter of the driving field, reading $\varphi$ for $t\in(0,\tau]$, $-\varphi$ for $t\in(\tau,2\tau]$, $-\varphi+\pi$ for $t\in(2\tau,3\tau]$, and $\varphi+\pi$ for $t\in(3\tau,4\tau]$.
With phase parameters quenched in this manner, one has $H_{k}(t)=\Omega(t)(\cos\varphi\sigma_{x}+\sin\varphi\sigma_{y})$. That is, the effective Hamiltonians become the common Hamiltonian $\Omega(t)(\cos\varphi\sigma_{x}+\sin\varphi\sigma_{y})$ over all time intervals.
If we further require
$\int^{4\tau}_{0}\Omega(t)dt=\theta/2$,
Eq. (\ref{eq2}) indicates
\begin{align}\label{eq3}
\mathcal{U}=e^{-i\theta(\cos\varphi\sigma_{x}+\sin\varphi\sigma_{y})/2}\otimes
e^{-i4H_{E}\tau}+\mathcal{O}(\tau^{2}).
\end{align}
As such, to the first order of $\tau$, the system-environment coupling is eliminated and yet we still implement a gate operation, namely, rotating the qubit along an axis in the $xy$ plane by angle $\theta$.  Remarkably, this is achieved without requiring that $H_{\rm S}(t)$ commutes with the DD operations.

We are now ready to engineer the time dependence of the qubit Hamiltonian for higher-order protection of quantum gates.  As an example, we consider the second-order protection by CDD.
Because the commutators between the qubit driving Hamiltonian and the system part in the system-environment Hamiltonian with the higher-order protection are simply the single-qubit Pauli operators that can be compensated by the PDD sequence alone, the second-order protection with CDD can be realized through utilizing the time evolution protected by the DD operations $\{\sigma_{0},\sigma_{1},\sigma_{2},\sigma_{3}\}$ as the basic building block and then nesting it with a second layer of DD operations $\{\sigma_{0},\sigma_{1},\sigma_{2},\sigma_{3}\}$.  In doing so, we examine how the time dependence of $H_{\rm S}(t)$ should be engineered. The evolution operator with two layers of DD operations is given by
\begin{align}\label{eq4}
\mathcal{U}=&\prod^{3}_{m=0}\sigma_{m}\left[\prod^{3}_{k=0}
\mathcal{T}e^{-i\int^{(k+1)\tau/4}_{k\tau/4}
\sigma_{k}H_{\rm S}(t)\sigma_{k}dt}\right]\sigma_{m}\otimes{e^{-i4H_{E}\tau}}
+\mathcal{O}(\tau^{3})
\notag\\
=&\prod^{3}_{m=0}\prod^{3}_{k=0}\mathcal{T}e^{-i\int^{(k+1)\tau/4}_{k\tau/4}
(\sigma_{m}\sigma_{k})H_{\rm S}(t)(\sigma_{k}\sigma_{m})dt}
\otimes{e^{-i4H_{E}\tau}}
+\mathcal{O}(\tau^{3}),
\end{align}
where all the terms in the product  should be arranged from right to left with increasing $k$ or $m$.
We can analogously define the function appearing in the exponential as the effective Hamiltonian $H_{mk}(t)\equiv(\sigma_{m}\sigma_{k})H_{\rm S}(t)(\sigma_{k}\sigma_{m})$.
Taking the driving Hamiltonian as $H_{\rm{S}}(t)=\Omega_{x}(t)\sigma_{x}+\Omega_{y}(t)\sigma_{y}$ again, we can easily obtain the explicit form of the effective Hamiltonian.
Just like the previous example showcasing first-order protection, here we can
also quench the phase $\varphi$ without changing the Rabi frequency $\Omega(t)$ during time intervals as well as subintervals such that $H_{mk}(t)=\Omega(t)(\cos\varphi\sigma_{x}+\sin\varphi\sigma_{y})$.
With $\int^{4\tau}_{0}\Omega(t)dt=\theta/2$, we have
\begin{align}
\mathcal{U}=e^{-i\theta(\cos\varphi\sigma_{x}+\sin\varphi\sigma_{y})/2}\otimes
e^{-i4H_{E}\tau}+\mathcal{O}(\tau^{3}),
\end{align}
indicating that our one-qubit gate with second-order protection is achieved.  It is straightforward to extend our approach to cases of higher-order protection of quantum gates by increasing the concatenation level of CDD.

Let us now turn to more efficient higher-order protection of quantum gates. According to the UDD theory proposed for quantum memory protection, the pure dephasing term $\sigma_{z}\otimes{E_{z}}$ can be filtered by applying the nonequidistant decoupling operator $\sigma_{x}$. Likewise, the longitudinal relaxation term $\sigma_{x}\otimes{E_{x}}+\sigma_{y}\otimes{E_{y}}$ can be suppressed by applying $\sigma_{z}$ operations at special UDD timings \cite{Uhrig,Yang,West2010,Wang2011}. Building upon this, to average out general system-environmental interaction $H_{I}=\sigma_{x}\otimes{E_{x}}+\sigma_{y}\otimes{E_{y}}+\sigma_{z}\otimes{E_{z}}$, we  apply $n$ decoupling operations $\sigma_{x}$ to the quantum system at times $t_{j}=4\tau\sin^{2}[j\pi/(2n+2)]$, and during each time interval $[t_{j-1},t_{j}]$, we further apply $n$ decoupling operations $\sigma_{z}$ at times $t^{j}_{k}=t_{j}\sin^{2}[k\pi/(2n + 2)]$, where $j,\ k=1,...,n$ and $n$ is taken as an even number to facilitate our engineering approach.
Such a nested UDD protocol yields the following time evolution operator:
\begin{align}
\mathcal{U}=&\mathcal{U}({4\tau,t_{n}})\prod^{n}_{j=1}\sigma_{x}\mathcal{U}(t_{j},t_{j-1})
\notag\\
=&\prod^{n+1}_{j=1}\sigma_{x}^{j-1}\mathcal{U}(t_{j},t_{j-1})\sigma_{x}^{j-1},
\end{align}
where $t_{0}\equiv0$, $t_{n+1}\equiv4\tau$, all the terms in the product from right to left are arranged to have an increasing $j$, and $\mathcal{U}(t_{j},t_{j-1})$, the unitary evolution operator associated with the interval $[t_{j-1},t_{j}]$, is expressed by a product of subinterval evolution operators $\tilde{\mathcal{U}}(t^{j}_{k},t^{j}_{k-1})$  as
\begin{align}
\mathcal{U}(t_{j},t_{j-1})=&\tilde{\mathcal{U}}({t_{j},t^{j}_{n}})\prod^{n}_{k=1}\sigma_{z}
\tilde{\mathcal{U}}(t^{j}_{k},t^{j}_{k-1})
\notag\\
=&\prod^{n+1}_{k=1}\sigma_{z}^{k-1}\tilde{\mathcal{U}}(t^{j}_{k},t^{j}_{k-1})\sigma_{z}^{k-1},
\end{align}
with $t^{j}_{0}\equiv t_{j-1}$ and $t^{j}_{n+1}\equiv t_{j}$, and all the terms in the product from right to left are arranged to have an increasing $k$.
The UDD theory can guarantee that the effect of system-environment interaction is suppressed up to the $n$th order. That is,
\begin{align}
\mathcal{U}=&\prod^{n+1}_{j=1}\prod^{n+1}_{k=1}
\mathcal{T}e^{-i\int^{t^{j}_{k}}_{t^{j}_{k-1}}\left(\sigma^{j-1}_{x}\sigma^{k-1}_{z}\right)
H_{\rm S}(t)\left(\sigma^{k-1}_{z}\sigma^{j-1}_{x}\right)dt}\otimes{e}^{-i4H_{E}\tau}
\notag\\
&+\mathcal{O}(\tau^{n+1}).
\end{align}
To find out how to engineer the time dependence of the driving Hamiltonian $H_{\rm S}(t)$, we define effective Hamiltonians $H^{j}_{k}(t)\equiv(\sigma^{j-1}_{x}\sigma^{k-1}_{z})
H_{\rm S}(t)(\sigma^{k-1}_{z}\sigma^{j-1}_{x})$, i.e., those terms appearing in the exponentials in the equation above.
Using the same driving Hamiltonian introduced above, one obtains the effective Hamiltonian as
$H^{j}_{k}(t)=(-1)^{k-1}\Omega_{x}(t)\sigma_{x}+(-1)^{k+j}\Omega_{y}(t)\sigma_{y}$, explicitly depicting the impact of our nested UDD operations on the system Hamiltonian.  To combat against these unwanted changes to the system Hamiltonian, we can just adjust only the phase parameter of the qubit driving field, such that for $t\in[t^{j}_{k-1},t^{j}_{k}]$,  $\Omega_{x}(t)=(-1)^{k-1}\Omega(t)\cos\varphi$, and $\Omega_{y}(t)=(-1)^{k+j}\Omega(t)\sin\varphi$. The evident outcome of this phase-quenching approach is that all the effective Hamiltonians will be the same, namely, $H^{j}_{k}(t)=\Omega(t)(\cos\varphi\sigma_{x}+\sin\varphi\sigma_{y})$. Assuming $\int^{4\tau}_{0}\Omega(t)dt=\theta/2$, then one arrives at
\begin{align}
\mathcal{U}=e^{-i\theta(\cos\varphi\sigma_{x}+\sin\varphi\sigma_{y})/2}\otimes
e^{-i4H_{E}\tau}+\mathcal{O}(\tau^{n+1}),
\end{align}
which is a quantum gate protected by nested UDD to the $n$th order.

Our approach illustrated above  can be extended to the engineering of various qubit-qubit interactions for higher-order protection of two-qubit operations.
For instance, the higher-order protection of the cross-resonance (CR) gate
$U_{\mathrm{CR}_{\theta}}\equiv\exp(-i\theta\sigma_{z}\otimes\sigma_{x}/2)$ can be implemented through engineering the following driving Hamiltonian:
\begin{align}\label{eq5}
H_{\rm S}=J(t)\sigma_{z}\otimes\sigma_{x},
\end{align}
where $J(t)$ is the coupling parameter between two qubits.
The CR gate is one of the typical two-qubit gates widely used in superconducting qubits with only microwave control \cite{Krantz}. It can be used to generate the CNOT gate by combining only two single-qubit gates such that $U_{\mathrm{CNOT}}=\exp(-i\pi\sigma_{z}/4)\otimes\exp(-i\pi\sigma_{x}/4)U_{\mathrm{CR}_{-\pi/2}}$ up to an unimportant global phase $\pi/4$.
In the aforementioned second-order protection of one-qubit gates with CDD, we conclude that the commutators between the driving Hamiltonian and the system part in the system-environment Hamiltonian are just the single-qubit Pauli operators and thus can be compensated by the nested PDD sequence alone. However, for higher-order protection of two-qubit gates, the situation is different
because the system part in the system-environment interaction not only includes single-qubit operators $\sigma_{x}$, $\sigma_{y}$ and $\sigma_{z}$ (i.e., $\sigma_{x}\otimes{I}$, $\sigma_{y}\otimes{I}$ and $\sigma_{z}\otimes{I}$ on the first qubit and ${I}\otimes\sigma_{x}$, ${I}\otimes\sigma_{y}$ and ${I}\otimes\sigma_{z}$ on the second qubit), but also contains the
combinations of operators acting on different qubits (as induced by the commutation relations between two-qubit coupling terms of the system Hamiltonian and the system part in the system-environment Hamiltonian).  The resulting combined operators are now $\sigma_{x}\otimes\sigma_{y}$, $\sigma_{x}\otimes\sigma_{z}$, $\sigma_{y}\otimes\sigma_{x}$, $\sigma_{y}\otimes\sigma_{z}$, $\sigma_{z}\otimes\sigma_{x}$ and $\sigma_{z}\otimes\sigma_{y}$ along with $\sigma_{x}\otimes\sigma_{x}$, $\sigma_{y}\otimes\sigma_{y}$ and $\sigma_{z}\otimes\sigma_{z}$.
For the second-order protection with CDD, if we analogously exploit the time evolution protected by PDD operations $\{\sigma^{\otimes2}_{0},\sigma^{\otimes2}_{1},\sigma^{\otimes2}_{2},\sigma^{\otimes2}_{3}\}$ as a basic building block and then nest it into a second layer of the same DD operations,
it is clear that the first six terms of these listed two-qubit operators along with the six single-qubit operators in the interaction Hamiltonian can be eliminated,  whereas the last three two-qubit terms will be intact. Therefore, most terms ($12/15=80\%$) in the interaction Hamiltonian can be eliminated using our nested PDD sequence with the effective Hamiltonians given by $H_{mk}(t)\equiv(\sigma_{m}\sigma_{k})^{\otimes2}H_{\rm S}(t)(\sigma_{k}\sigma_{m})^{\otimes2}$.
Using the effective Hamiltonians, it can be seen that
the impact of DD pulses on the system Hamiltonian in Eq.~(\ref{eq5}) will flip the sign of the qubit-qubit interaction in the second and fourth subintervals of the first and third intervals and the first and third subintervals of the second and the fourth intervals. In other words, we can just reverse the direction of the driving field in the above subintervals to cancel the unwanted action of the CDD pulses.
As a result of such engineering, the target CR gate $U_{\mathrm{CR}_{\theta}}$ can be obtained with the requirement $\int^{4\tau}_{0}J(t)dt=\theta/2$.
For the other error operators, namely,  $\sigma_{x}\otimes\sigma_{x}, \sigma_{y}\otimes\sigma_{y}, \sigma_{z}\otimes\sigma_{z}$ in the system-environment interaction Hamiltonian after applying the aforementioned DD sequence, we can further apply the DD operator $\sigma_{z}\otimes{I}$ to eliminate the two terms $\sigma_{x}\otimes\sigma_{x}, \sigma_{y}\otimes\sigma_{y}$ and then nest it into the next layer of the DD operator $I\otimes\sigma_{x}$ to eliminate the term $\sigma_{z}\otimes\sigma_{z}$. Certainly, the engineering of the time dependence of the gate Hamiltonian can be done analogously according to the corresponding effective Hamiltonian.

\section{Numerical experiments}

Our higher-order protection promises to retain a high fidelity even for a slow one-qubit gate. For the two-qubit gate, the higher-order protection promises enhanced fidelity using single-qubit gates as DD operations. In the following, we demonstrate these through numerical experiments, with one particular focus on the impact of the finite duration of the DD operations on the performance of gate protection.

Let us now use some numerical simulations as proof-of-concept experiments, by considering a qubit system interacting with a five-spin bath. The interaction between the qubit and each of the bath spins physically originates from an isotropic Heisenberg coupling $\sigma_{x}^{s}\sigma_{x}^{b}+\sigma_{y}^{s}\sigma_{y}^{b}+\sigma_{z}^{s}\sigma_{z}^{b}$ and an antisymmetric Dzyaloshinskii-Moriya coupling $\sigma_{x}^{s}\sigma_{y}^{b}-\sigma_{y}^{s}\sigma_{x}^{b}$ with a common coupling strength parameter $\epsilon$ for convenience \cite{Mousolou,Doherty}.
Here, the first qubit represents the system spin and and all other spins represent the bath.
Though simple, such a model construction captures the essence that the qubit system is subject to both dephasing and population relaxation and hence suffices to illustrate how our engineering approach leads to some higher-order protection. The spin-bath Hamiltonian adopted above may also model some residue interaction between a qubit system under gate operation and its surrounding qubits.
The performance of higher-order protection is characterized by the fidelity
$F=\bra{\phi}\rho\ket{\phi}$, where $\ket{\phi}$ is a target output state and $\rho$ is a real output state.

\begin{figure}[t]
  \includegraphics[scale=0.41]{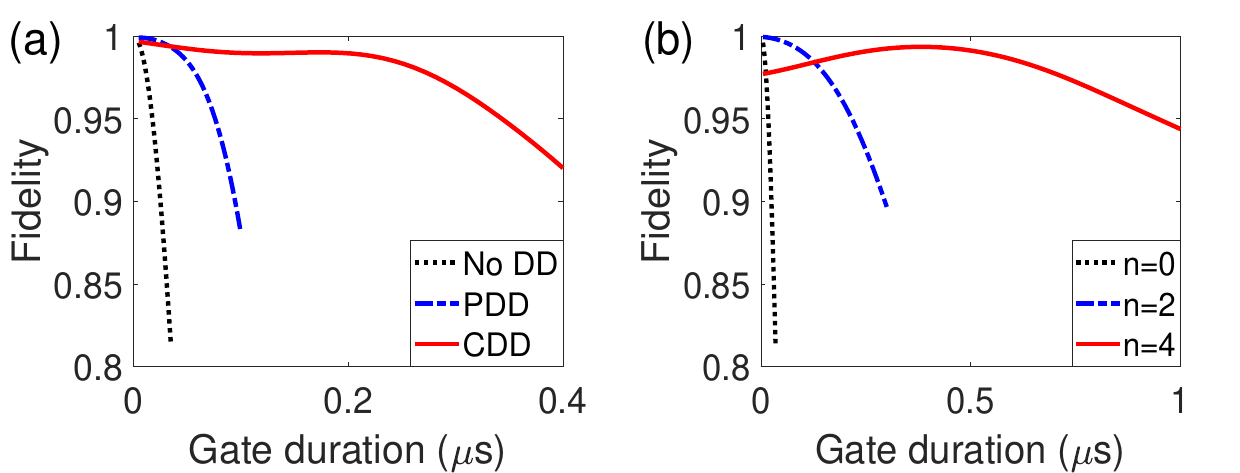}
  \caption{The fidelity of a one-qubit gate changes with increasing the gate duration. (a) The fidelities under PDD with the first-order protection (blue) and CDD with the second-order protection (red), and without DD protection (black). (b) The fidelities under nested UDD protection with the orders $n=2$ (blue) and $4$ (red), and without DD protection (black).}
  \label{Fig1}
\end{figure}

For one-qubit gates, we set the initial state as $(\ket{0}+\ket{1})/\sqrt{2}$, the ideal gate as $U=(\sigma_{x}+\sigma_{y})/\sqrt{2}$ [the corresponding gate parameters are $\theta=\pi$ and $\varphi=\pi/4$, see Eq.~(\ref{eq3})], and the system-bath coupling strength as $\epsilon=2\pi\times1\mathrm{MHz}$.
Unless specified otherwise, the strength of our square-shaped DD pulse is chosen to be $2\pi\times100\ \mathrm{MHz}$.
This means that we utilize the finite-time gate operations rather than the ideal instantaneous-pulse operations as our actual DD pulses.
The DD protection of quantum gates is then performed by nesting the finite-time DD operations between the piecewise engineered time evolution according to the outlined approach.
To illustrate how our higher-order protection can achieve a high fidelity of one-qubit gates over
a long time duration, we tune the gate duration $4\tau$ over a wide range.
In Fig.~\ref{Fig1}(a), we present the fidelity $F$ versus the gate duration under the PDD and CDD with first-order and second-order protections, represented  by the blue and red lines.
As a comparison, we also plot the gate fidelity without any DD protection, as shown by the black line in the same figure.
The results in Fig.~\ref{Fig1}(a) show that, while the first-order protection can already tolerate  a longer gate duration than that of the bare gate, the second-order scheme offers a much better performance.  For example, the first-order and second-order schemes yield gate fidelity $>95\%$ for  gate duration $0.0125~\mathrm{\mu s}$ and $0.08~\mathrm{\mu s}$, respectively, both being much longer than what the bare gate can last $0.007~\mathrm{\mu s}$ at the same fidelity level.  These results are encouraging because the applied DD pulses here have a finite width.

Next we investigate the performance of nested UDD in protecting the one-qubit gate using our protocol outlined above. Figure~\ref{Fig1}(b) presents the gate fidelity with an increasing gate duration for second-order and fourth-order UDD, as shown by the blue and red lines, respectively (the gate fidelity without DD is also represented by the black line as a reference).  There it is seen that the second-order and fourth-order UDD protection retains the gate fidelity $>90\%$ for  gate duration $0.07~\mathrm{\mu s}$ and $0.2~\mathrm{\mu s}$, respectively, about $7$ and $20$ times longer than the duration $0.01~\mathrm{\mu s}$ of a bare gate at the same fidelity level. It is interesting to note that such a great enhancement of gate duration at a high fidelity level using higher-order UDD protection is different. Of course,  the numerical findings here do not represent at all what one can achieve on an actual quantum computing platform, but still indicate that our higher-order protection approach of gate operations is indeed promising.

\begin{figure}[t]
\includegraphics[scale=0.41]{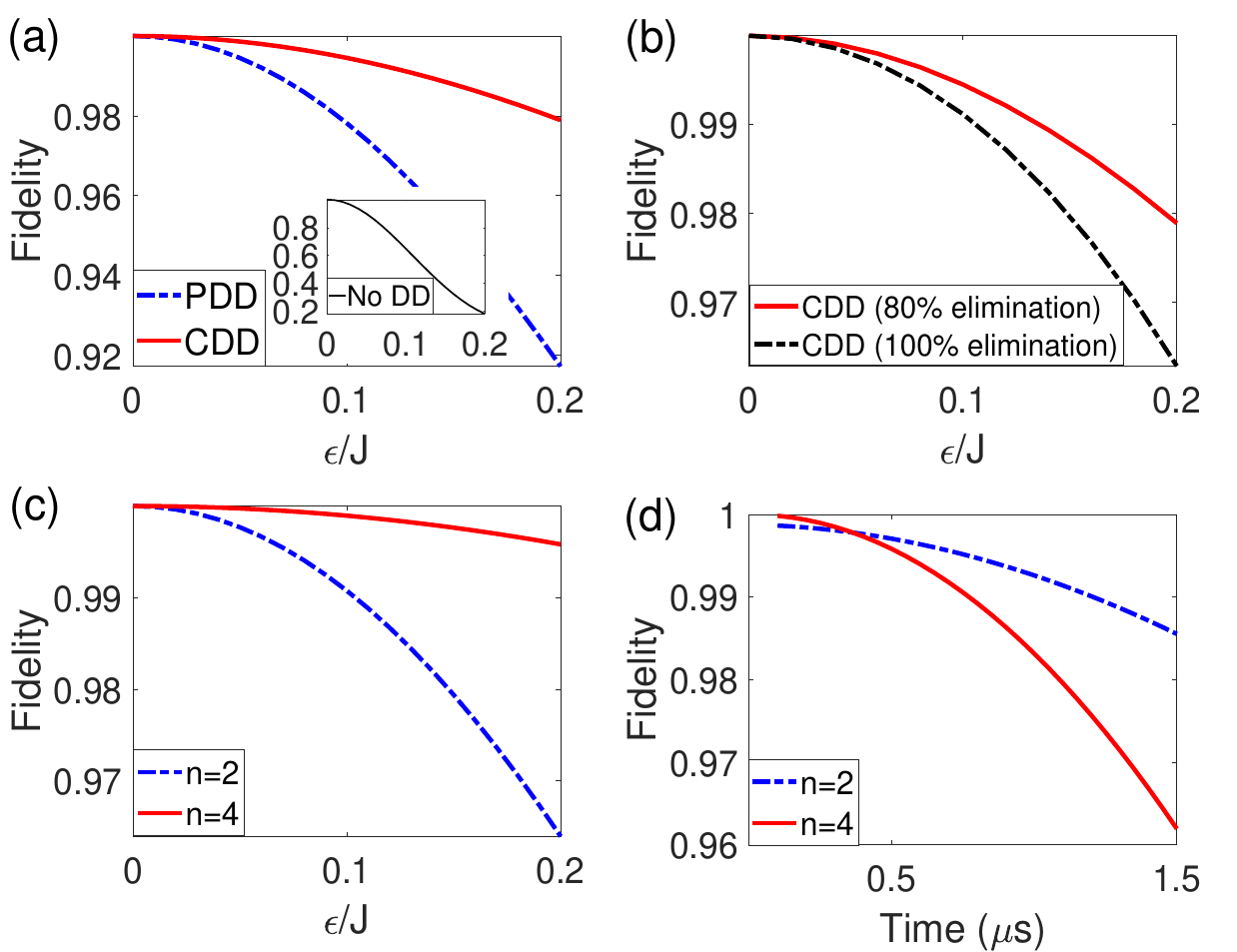}
\caption{The fidelity of a two-qubit gate using single-qubit operations as DD pulses. (a) The gate fidelity of PDD with the first-order protection (blue) and CDD with the second-order protection (red) versus the system-environment coupling strength over $\epsilon\in[0,0.2J]$, where the gate fidelity without protection is also presented in the small block as a comparison. In both cases, the duration of one single DD pulse is chosen to be $2.5~\mathrm{ns}$. (b) The gate fidelity for CDD protection when eliminating most of the terms with two layers (red) and eliminating all the terms with four layers (black). (c) The gate fidelity under nested UDD protection with the orders of $n=2$ (blue) and $4$ (red) for $\epsilon\in[0,0.2J]$, where the duration of one DD pulse is chosen as $0.5~\mathrm{ns}$ for $n=4$ and $2.5~\mathrm{ns}$ for $n=2$. (d) The gate fidelity vs the time duration of one DD pulse, namely,  $\tau_{\rm DD}\in[0.1~\mathrm{ns},1.5~\mathrm{ns}]$ for $n=2$ (blue) and $n=4$ (red) with $\epsilon=0.2J$.}
  \label{Fig2}
\end{figure}

Finally, we turn to a two-qubit gate case.  The ideal gate under consideration is
$U_{\mathrm{CR}_{-\pi/2}}=\exp(i\pi\sigma_{z}\otimes\sigma_{x}/4)$, the coupling parameter between two qubits is $J=2\pi\times5\mathrm{MHz}$ [see Eq.~(\ref{eq5})], and the initial state is $\ket{11}$. In Fig.~\ref{Fig2}(a), we present the fidelity $F$ versus the system-environment coupling strength over $\epsilon\in[0,0.2J]$ under PDD (first-order) and CDD (second-order), represented by the blue and red lines. The gate fidelity without any DD protection is also plotted as a small block in the same figure as a comparison.
The results show that both first-order and second-order protocols yield gate fidelities much better than that of the bare gate. For example, for the case $\epsilon=0.2J$, i.e., the system-environment coupling strength is $0.2$ times of the gate Hamiltonian strength such that the fidelity of the bare gate is as low as $18.01\%$, the first-order protection of PDD can improve the fidelity to $91.73\%$ while the second-order protection of CDD can boost the fidelity back to $97.89\%$. It is worth emphasizing that for the CDD protection, if we eliminate all the second-order terms in the interaction Hamiltonian, we need to concatenate four layers of the DD sequence, thereby  requiring a significant overhead of DD operations. However, if we eliminate most of the second-order terms ($80\%$ terms as discussed above), we only need two layers of the DD sequence and hence save a large number of DD operations. Considering that each individual DD gate operation also suffers from decoherence, there exists a trade off between the number of error terms to be eliminated and the layers of DD sequence we consider to apply.  As is shown in Fig.~\ref{Fig2}(b), we find that the CDD protection to eliminate most of the error terms with two layers yields a higher gate fidelity than that to eliminate all the terms with four layers. For this consideration, we choose the CDD protocol with two nested layers in the above discussion and recommend to adopt approaches with less layers for higher-order protections in the discussions below.

Our numerical experiments can also illustrate the performance of our approach when using nested UDD to protect the two-qubit gate to the $n$th order. Figure~\ref{Fig2}(c) presents the corresponding gate fidelity with $n=2$ and $4$ over a wide range $\epsilon\in[0,0.2J]$. The gate fidelity is seen to be greatly enhanced when increasing the order of protection. In the case of fourth-order protection, we need $24$ non-instantaneous DD operations to both qubits and hence the duration of each individual DD operation will be more consequential. The results in Fig.~\ref{Fig2}(c) are obtained with the duration of DD operation shortened by a factor of $5$ for the fourth-order case. The resultant gate fidelity with the second- and fourth-order protections can be as high as $96.39\%$ and $99.58\%$, respectively, even for a relatively large system-environment coupling strength $\epsilon=0.2J$. To investigate the influence of DD duration on our higher-order protection, we further plot the gate fidelity $F$ versus the duration of one single DD operation, namely, $\tau_{\mathrm{DD}}\in[0.1~\mathrm{ns},1.5~\mathrm{ns}]$.  The results are shown in Fig.~\ref{Fig2}(d), with the system-environment coupling strength $\epsilon=0.2J$. One sees that the fidelity of the fourth-order protection can reach $99.9\%$ whereas the fidelity of the second-order protection does not. Furthermore, the fidelities of the fourth-order and second-order protections intersect at $\tau_{\mathrm{DD}}=0.35~\mathrm{ns}$, indicating that in real life applications with finite-time DD pulses that are not sufficiently short, the second-order nested UDD can be better than the fourth-order version because the latter also induce extra decoherence due to the use of a large number of finite-duration DD pulses.

\section{Physical demonstration on an IBM quantum computer}

Having shown the feasibility of our higher-order protection of quantum gates through numerical experiments above, we shall now examine the actual performance of our approach on an IBM quantum computer.
However, because our access to IBM hardware does not allow us to insert single-qubit gates within one two-qubit gate, we consider higher-order protection of the combination of a series of CNOT gates instead.  To that end, we use the original QISKIT API~\cite{Abhari}
to realize the first- and second-order protections of the CNOT gate series under PDD and CDD.

\begin{figure}[t]
  \includegraphics[scale=0.14]{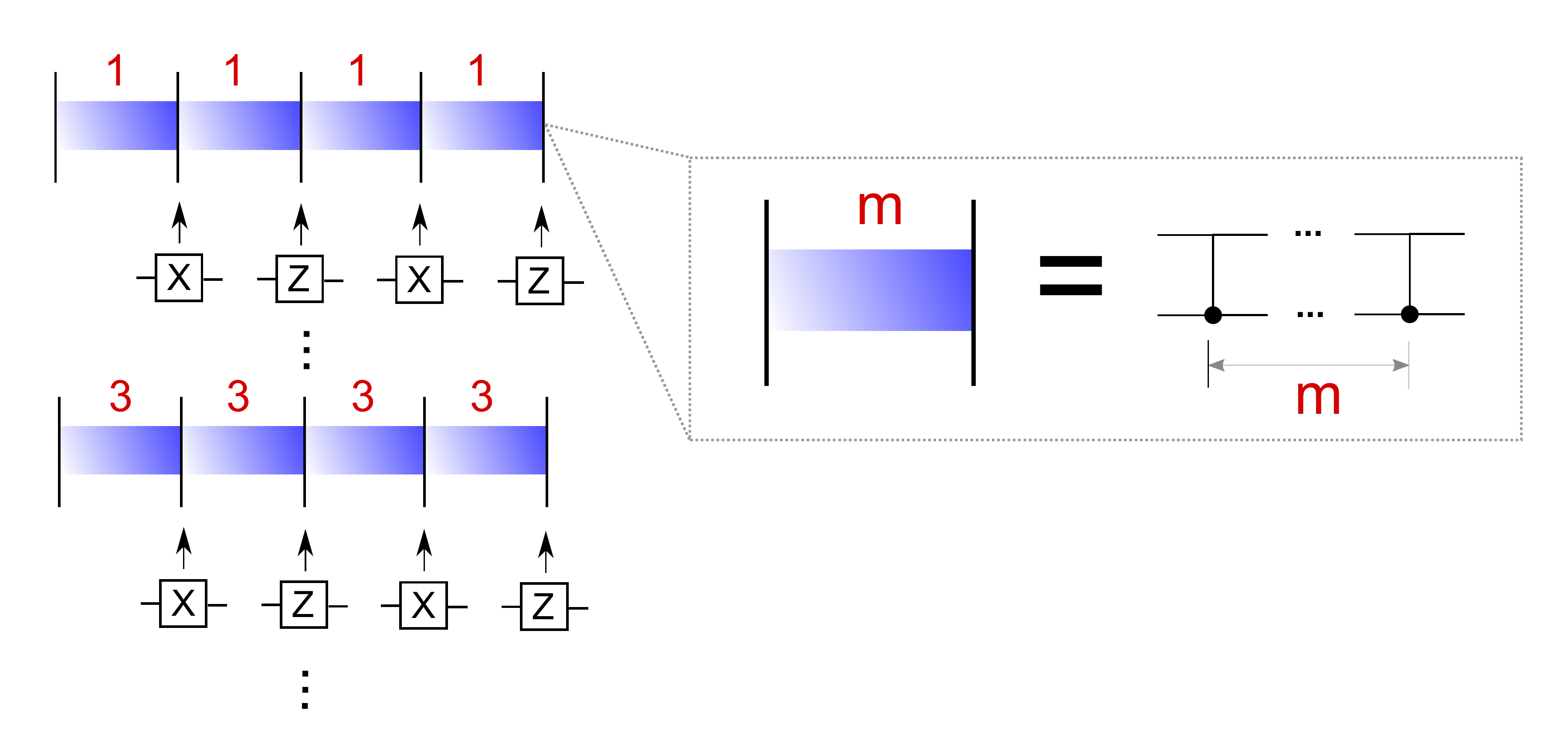}
  \caption{The schematic for implementing the first-order protection of CNOT operations on an IBM quantum processor. Four blocks of quantum gates are concatenated with each block containing
   $m$ gates. One-qubit gates are inserted according the PDD protocol between neighboring blocks to realize the first-order protection.}
  \label{Fig3}
\end{figure}

\begin{figure}[t]
  \includegraphics[scale=0.55]{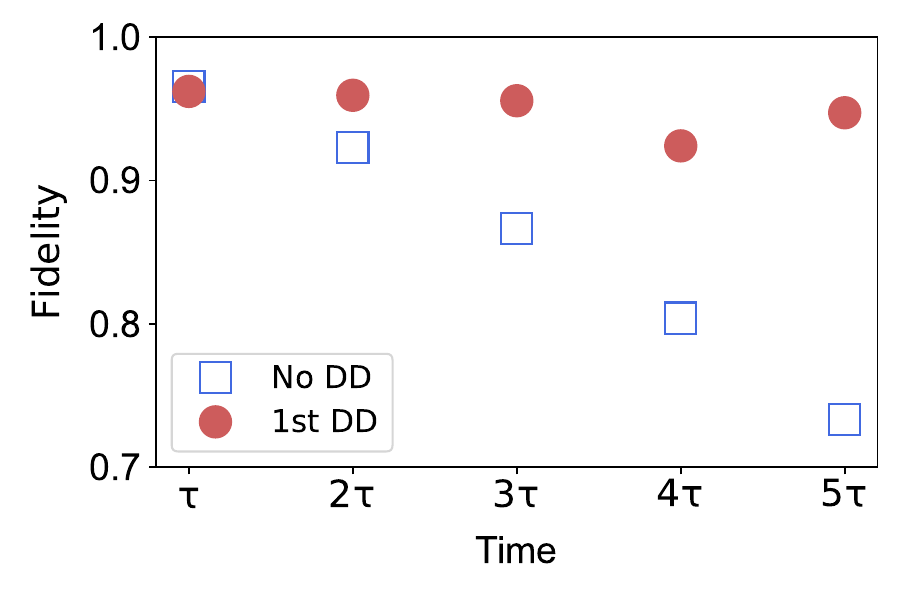}
  \caption{The overall gate fidelity of combined CNOT gates under the first-order protection with PDD and Hamiltonian engineering,  obtained from the IBM quantum processor. Here, the time unit $\tau=2.64 \times 10^2~\mathrm{ns}$ represents the duration to execute totally four CNOT gates (one in each block). Fildelity in the case of $5\tau$ is for a total of 20 combined CNOT gates.  The results are obtained from IBM\_KYOTO hardware, and the total shots for each data are 10000.}
  \label{Fig4}
\end{figure}

For the first-order protection, we concatenate four blocks of quantum gates with the number of gates in each block being $m\in\{1,2,3,4,5\}$. Figure~\ref{Fig3} shows two examples with $m=1$ and $m=3$.
We then insert the PDD sequence equidistantly between two neighboring blocks to implement the first-order protection. The PDD protocol elaborated in Eq.~(\ref{eq1}) requires us to sequentially insert $\sigma_x$ (denoted by X) and $\sigma_z$ (denoted by Z) single-qubit gates as the DD operations, also shown in Fig.~\ref{Fig3}.
Most importantly, according to our Hamiltonian engineering strategy elaborated above,  we can now coordinate the gate Hamiltonian to be applied in accord with the DD operations to be applied.  That is, to counteract the interference of the inserted DD operations on the time evolution, we apply pre-engineered two-qubit gates instead of the targeted CNOT gate. Specifically, if within each block there is  one CNOT gate, the four two-qubit gates to be applied to the four blocks are given by
\begin{align} \label{eq6}
U_{1}&=\ket{0}\bra{0}\otimes{I}+\ket{1}\bra{1}\otimes{\sigma_x},
\notag\\
U_{2}&=\ket{1}\bra{1}\otimes{I}+\ket{0}\bra{0}\otimes{\sigma_x},
\notag\\
U_{3}&=\ket{1}\bra{1}\otimes{I}-\ket{0}\bra{0}\otimes{\sigma_x},
\notag\\
U_{4}&=\ket{0}\bra{0}\otimes{I}-\ket{1}\bra{1}\otimes{\sigma_x}.
\end{align}
It is straightforward to verify that these deformed two-qubit gates will be transformed to the expected CNOT gate once accounting for the interference of the PDD protocol.
Interestingly, due to the explicit construction of the DD operation here, if within each block there is an even number of CNOT gates, then there is no need for such engineering because the overall action of these CNOT gates (which is identity) is not affected by the DD pulses.
This being the case, if within each block there is an odd number of CNOT gates,   we only need to engineer the first (or last) two-qubit gates in the $j$th block to be $U_{j}$ defined in Eq.~(\ref{eq6}).
With these arrangements, we examine the gate fidelity decay as a function of the duration to execute all the CNOT gates.  For that purpose, we let $\tau$ denote the duration to execute four CNOT gates, with one block accommodating one CNOT gate. The raw counts are obtained from the sampling of the quantum circuit's execution on the IBM quantum processor. The results are shown in Fig.~\ref{Fig4}, with the initial state taken as $\ket{00}$ and each circuit executed for $10000$ times. From the results, we can see that the first-order protection of the CNOT gates clearly offers an improvement of gate fidelity as compared with the unprotected case. This improvement becomes more obvious as the total number of executed CNOT gates increases.

\begin{figure}[t]
\includegraphics[scale=0.13]{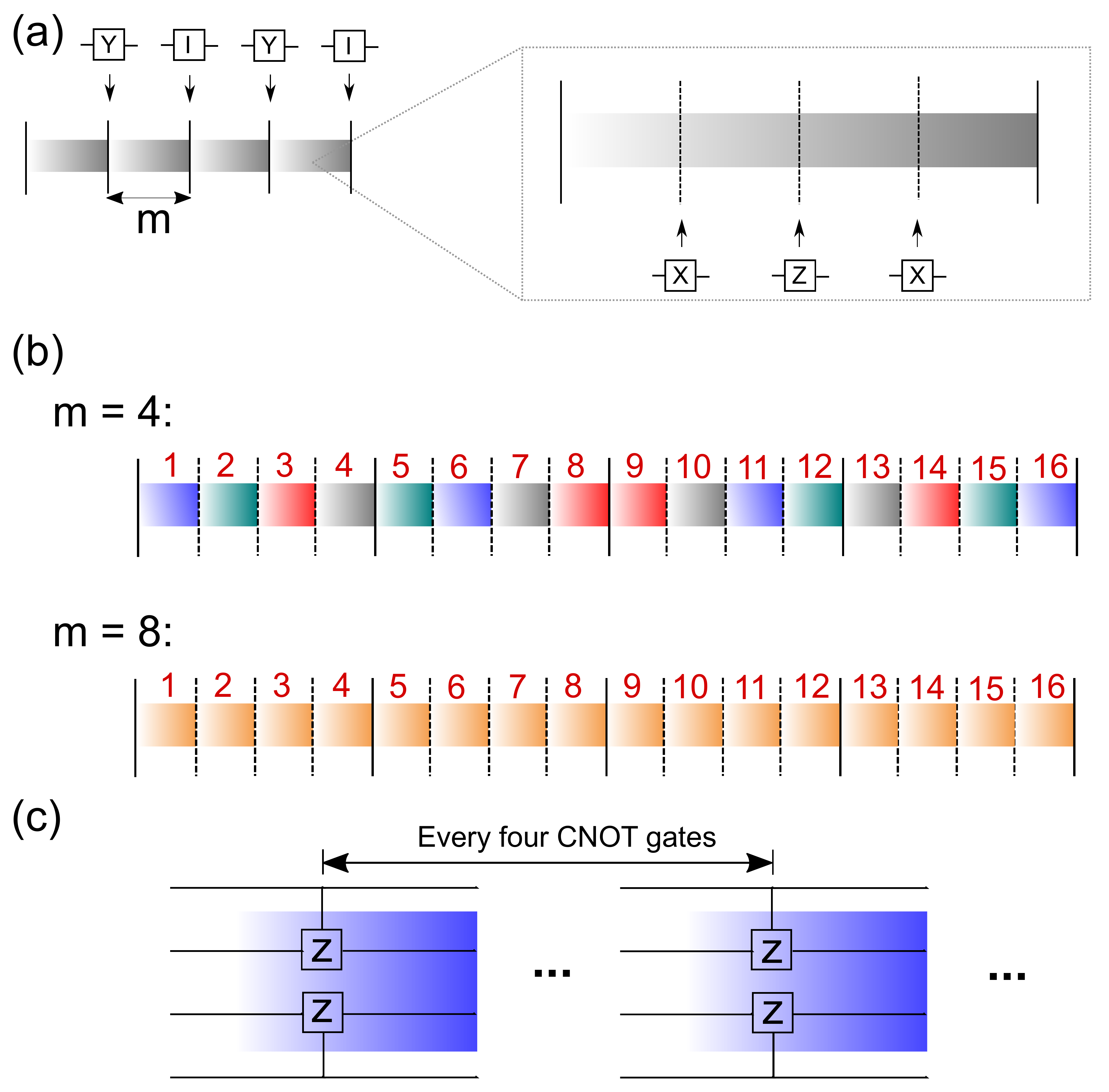}
\caption{The setup for implementing the second-order protection of CNOT gates on the IBM quantum processor, based on Hamiltonian engineering coordinated with CDD.
(a) Four blocks of quantum gates are concatenated to form a noisy time evolution and each block is further divided into four subblocks with each subblock containing several two-qubit gates. The CDD sequence is then inserted to implement the second-order protection.
(b) The engineered quantum gates when $m=4$ and $8$. For $m=4$, the first four engineered gates $U_{1}$, $U_{2}$, $U_{3}$, and $U_{4}$ are represented by the first four colors, and the other 12 engineered gates chosen from $U_{1}$, $U_{2}$, $U_{3}$, and $U_{4}$ are depicted using different colors.
For $m=8$, each color represents a CNOT gate.
(c) Crosstalk between the data qubits and surrounding qubits is introduced by adding two surrounding qubits to the two-qubit quantum circuit along with a controlled-$Z$ gate between the data qubit and surrounding qubit after every four CNOT gates.}
\label{Fig5}
\end{figure}

\begin{figure}[t]
  \includegraphics[scale=0.56]{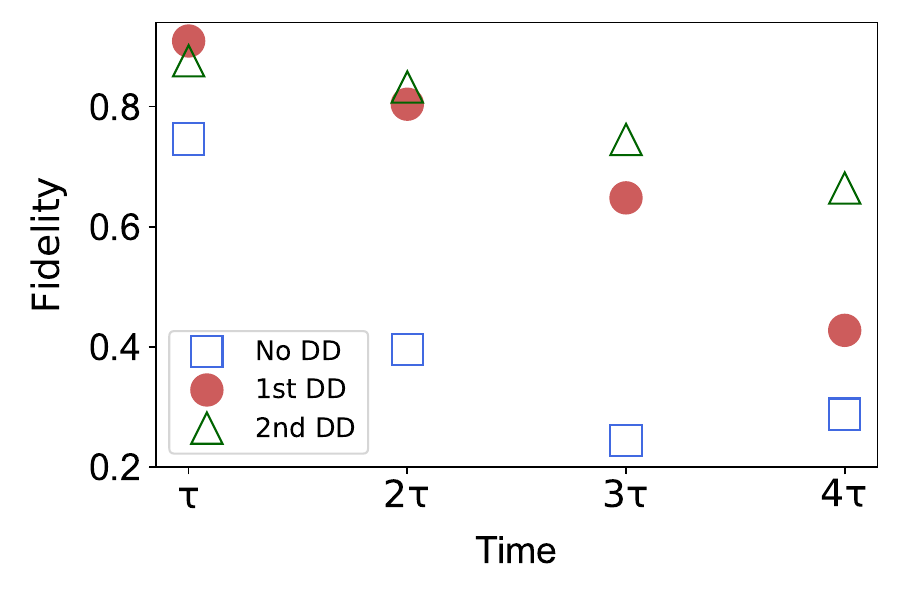}
  \caption{The overall gate fidelity of combined CNOT gates under second-order protection on the IBM quantum processor, as compared with the corresponding results under first-order protection and no protection. Here, $\tau=1.06\times10^3~\mathrm{ns}$ represents the duration to execute 16 CNOT gates (one CNOT gate in each subblock).  Fildelity in the case of $4\tau$ is for a total of 64 combined CNOT gates (four CNOT gates in each subblock).    Note that the results for the first-order DD protection presented here are not achieved at the same time as those presented in Fig.~\ref{Fig4}, though obtained from the same IBM$\_$KYOTO hardware. The total shots for each data are 10000.}
  \label{Fig6}
\end{figure}

Let us now turn to the physical demonstration of the second-order protection of the CNOT gate series based on Hamiltonian engineering coordinated with the CDD protocol. To this end, we still concatenate four blocks of quantum gates, but in each block we stack two-qubit gates in the number of $m\in\{4,8,12,16\}$, as is shown in Fig.~\ref{Fig5}(a).
Furthermore, we divide each block into four subblocks and hence there are 16 subblocks in total.
Accordingly, if $m=4$, then each subblock accommodates only one two-qubit gate; whereas for $m=8$, each subblock accommodates two two-qubit gates.
We then insert the CDD sequence between the blocks and between the subblocks to implement the second-order protection. The actual one-qubit gates applied as DD operations at two different layers are indicated in Fig.~\ref{Fig5}(a).
To counteract the interference of the CDD sequence on the targeted time evolution, we can now
apply engineered gates coordinated with CDD, analogous to the above first-order case.  In particular, for $m=4$, the first four two-qubit gates in the first block are precisely the above-defined $U_{1}$, $U_{2}$, $U_{3}$, and $U_{4}$ in Eq.~(\ref{eq6}), followed by the other 12 two-qubit gates chosen from $U_{1}$, $U_{2}$, $U_{3}$, and $U_{4}$.  The explicit sequence of all the 16 two-qubit gates is depicted in  Fig.~\ref{Fig5}(b) using different colors.  It is the overall action of these two-qubit gates plus the DD operations that may protect the CNOT gate series to the second order.  The same sequence of two-qubit gates will be applied if $m=4\times(2r-1)$, where $r$ is a positive integer. If $m=4\times2r$, i.e., each subblock contains an even number of CNOT gates, then their overall action within each subblock commutes with the DD operations, and hence
there is no need to deform the CNOT gates, as also indicated in Fig.~\ref{Fig5}(b).
To demonstrate that our Hamiltonian engineering approach of higher-order protection can operate in an integrated quantum circuit, we also deliberately introduce the crosstalk between the data qubits and some surrounding qubits. This is done by applying a controlled-$Z$ gate to the data and surrounding qubits after every four CNOT gates, shown in Fig.~\ref{Fig5}(c).
Figure~\ref{Fig6} presents the overall fidelity of the combined CNOT gates under second-order protection versus the total time duration to execute the CNOT gates. For comparison, we also present the corresponding results obtained under the first-order protection and without any protection in the same figure.  It is seen that the second-order protection clearly offers an enhanced fidelity compared with what can be achieved with bare CNOT gates.  Furthermore, the fidelity with second-order protection is slightly lower than that with first-order protection in the case of 16 CNOT gates. However, as the total number of CNOT gates increases, the case with second-order protection wins over the case with first-order protection.

\section{Discussions}

The mitigation of noise and decoherence effects on the performance of quantum gates is a challenging optimization problem.    This line of research, even at the the one-qubit level \cite{Zeng,Dong,Yi},  continues to  advance quantum computing and quantum simulation technologies \cite{Montangero,Rebentrost,Tsai,Calzavara,Huang,Larrouy,Motzoi,Hailski, Motzoi2}.
Machine learning techniques will play more important roles to realize robust quantum computing with optimized performance \cite{Zahedinejad,Niu,Fillingham}.  To guide future machine learning based optimization, some general principle towards higher-order protection of quantum gates would be extremely valuable in finding the optimal design of quantum gate Hamiltonians.   In this work, we showed that highly efficient schemes such as CDD and UDD for quantum memory protection can be used for the protection of quantum gates at different orders, provided that there is coordination between the DD operations and the time dependence of the gate Hamiltonians.   To place a two-qubit gate under high-order protection,  the basic building block may necessarily involve a number of high-fidelity single-qubit gates and the time dependence of the applied two-qubit gate Hamiltonian should be engineered accordingly.

This work has thus offered a promising starting solution towards optimized designs of robust quantum gates.    Throughout, we only assume that the environment noise spectrum has a hard cutoff so that the higher-order schemes for quantum memory protection can work well (DD against an environment with a soft noise spectrum cutoff remains a challenge \cite{Liu2013}).
If there is more specific spectral information about the environment noises, then the time intervals between DD pulses can be further optimized according to the noise spectrum \cite{Gong11} and as a result, quantum gate protection using our concept can directly benefit from such optimization that can further drastically improve the performance of decoherence suppression. Our Hamiltonian engineering approach in coordination with DD hence makes the measurement of environment spectrum much more relevant than before to the realization of high-fidelity quantum gates.

Through numerical experiments using square-shaped pulses and physical demonstration on an IBM quantum processor, we demonstrated that our approach is effective with noninstantaneous DD operations and hence experimentally feasible on noisy intermediate scale quantum computers. If pulse shaping is further introduced to the DD pulses and the driving Hamiltonians, then we foresee that the higher-order protection scheme proposed in this work may further improve its performance. The long-term hope is that quantum gates protected under higher-order protection protocols can eventually provide a versatile toolbox to noisy intermediate scale quantum computers in the actual mitigation of decoherence, where a large number of gate operations will be applied in an integrated and complex quantum circuit.

\begin{acknowledgments}
J.G. is grateful to Lorenza Viola for technical discussions and for her highly constructive comments on the first version of our manuscript. T.C. is grateful to Vinay Tripathi and Dario Poletti for fruitful discussions. We acknowledge the use of IBM Quantum services for this work via Singapore Quantum Engineering Programme (QEP). We also acknowledge the use of IBM Quantum Credits for this work. The views expressed are those of the authors, and do not reflect the official policy or position of IBM or the IBM Quantum team. This work was supported by the National Research Foundation, Singapore and A*STAR under its CQT Bridging Grant.
\end{acknowledgments}

\appendix
\section{Details of digital simulations on the IBM quantum processor}

\subsection{Qubits and quantum gates error}

For all the demonstration results presented in the main text, we executed the quantum circuits on the IBM quantum processor \cite{IBMQuantum}, which is a state-of-the-art quantum computing platform that leverages superconducting qubits. All the quantum circuits are constructed via QISKIT API, and executed via the QISKIT RUNTIME service \cite{Abhari}.

\begin{figure*}[t]
  \includegraphics[width=2 \columnwidth]{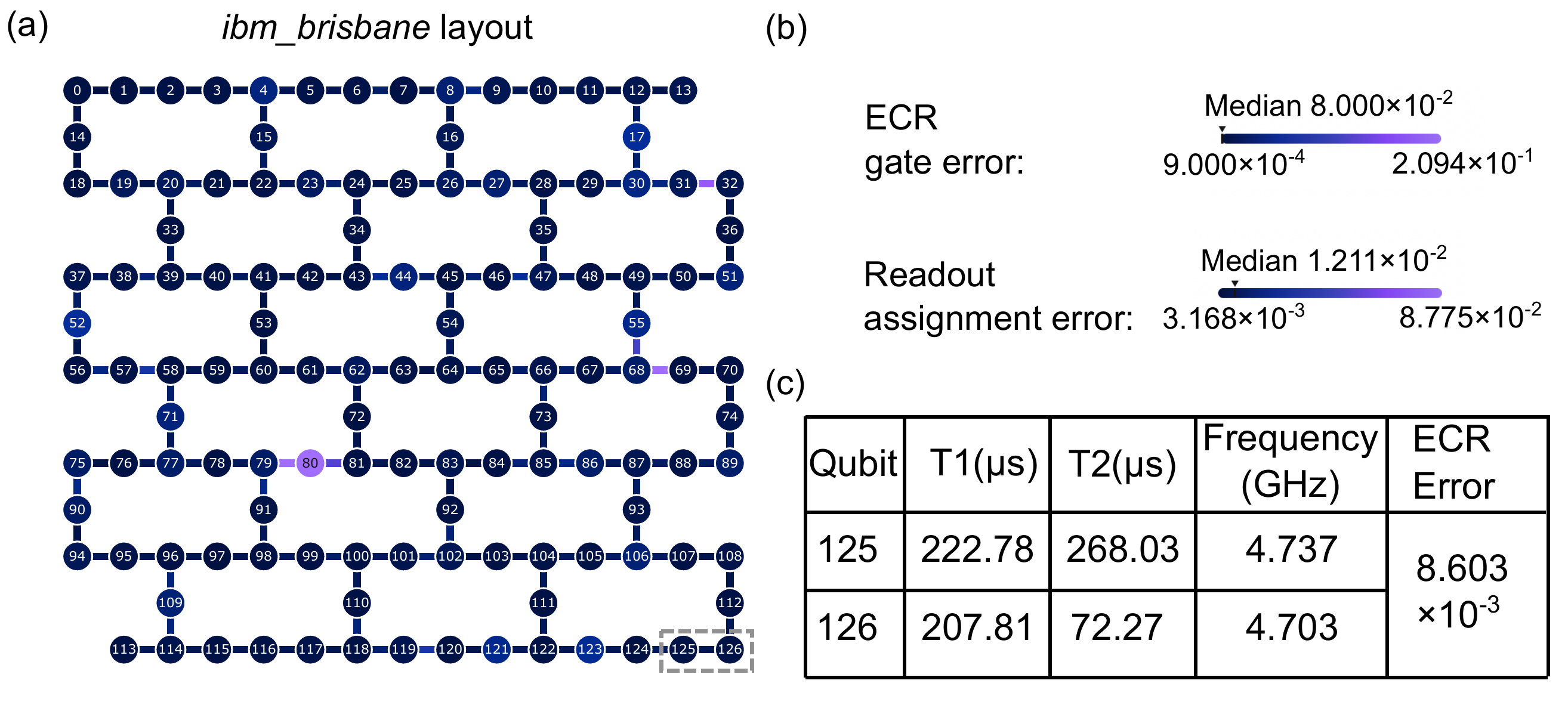}
  \caption{Details of the IBM quantum processor: (a) the geometric layout of IBM\_BRISBANE device of the 127-qubit Eagle type from IBM Q \cite{IBMQuantum}. The solid circles represent one single qubit, and each qubit is connected by a solid bond. The color for both the circle and the bond shows the amplitude of different types of gate error, i.e., the lighter the color is, the larger the error is. The dashed gray square indicates the two qubits and the connection bond chosen for implementing the algoirithm in the main text. (b) Averaged and two-qubit ECR gate error and the readout assignment error (measurement error). (c) Details of the 125th and 126th qubits' qualities: $T1$, $T2$, qubit frequency and the ECR gate error corresponding to their bond.}
  \label{Fig7}
\end{figure*}

The details of the IBM quantum processor, the quantum gates error as well as the qubit quality information are listed in Fig.~\ref{Fig7}. The layout of the IBM quantum processor ``IBM\_BRISBANE" is shown in Fig.~\ref{Fig7}(a), which consists of $127$ qubits in total. At the time, performing all the demonstration of the DD on IBM\_BRISBANE, the averaged two-qubit echoed cross-resonance (ECR) gate~\cite{Gambetta2020,Jurcevic2021} error as well as the readout assignment error (measurement error) is depicted in Fig.~\ref{Fig7}(b). The ECR gate is a native two-qubit entangled gate for the $127$-qubit IBM quantum processor. It achieves maximal entanglement and is functionally equivalent to a CNOT gate, subject to single-qubit pre-rotations. The echoing technique reduces unwanted terms (those not involving $ZX$) in an experimental setup. Specifically, this gate realizes:
\begin{align}
  \label{eq: ECR gate matrix}
  &{\rm ECR}=\frac{1}{\sqrt{2}}\begin{pmatrix}
    0 &1 &0 &i \\
    1 &0 &-i &0 \\
    0 &i &0 &1 \\
    -i &0 &1 &0
  \end{pmatrix}
\end{align}

Throughout this work, we choose the 125th and 126th qubits [gray dashed square in Fig.~\ref{Fig7}(a)] for the implementation of all quantum circuits, as the two-qubit ECR gate error shall not be too large [Fig.~\ref{Fig7}(c)] such that when stacking more two-qubit gates together, the error is moderate enough for our DD approach to work within the total circuit depth considered.

\subsection{Measurement and error mitigation}

The fidelity which was measured in the main text is defined as $F=|\langle \uparrow_a \uparrow_b|\psi \rangle|$, where the subscript $a$ and $b$ represent the target qubits for the stacked CNOT gates.

The error mitigation of the measurement error (readout assignment error) has already been included natively into the QISKIT RUNTIME option~\cite{Abhari}. Therefore, during the submission of the tasks, we set up the resilience level option to be $1$, and we find the difference between the error mitigation level $1$ and other levels is not large enough to affect the major results in this work.

Finally, we remark that for all results in this work, the total number of shots we applied is fixed at $100000$.


\begin{thebibliography}{99}

\bibitem{Knill} E. Knill, R. Laflamme, and W. H. Zurek, Resilient quantum computation, Science \textbf{279}, 342 (1998).
\bibitem{Preskill} J. Preskill, Reliable quantum computers, Proc. R. Soc. A \textbf{454}, 385 (1998).
\bibitem{Viola} L. Viola, E. Knill, and S. Lloyd, Dynamical decoupling of open quantum systems, Phys. Rev. Lett. \textbf{82}, 2417 (1999).
\bibitem{Viola1999} L. Viola, S. Lloyd, and E. Knill, Universal control of decoupled quantum systems, Phys. Rev. Lett. \textbf{83}, 4888 (1999).
\bibitem{Viola2003} L. Viola and E. Knill, Robust dynamical decoupling of quantum systems with bounded controls, Phys. Rev. Lett. \textbf{90}, 037901 (2003).
\bibitem{Khodjasteh} K. Khodjasteh and D. A. Lidar, Fault-tolerant quantum dynamical decoupling, Phys. Rev. Lett. \textbf{95}, 180501 (2005).
\bibitem{Uhrig} G. S. Uhrig, Keeping a quantum bit qlive by optimized $\pi$-pulse sequences, Phys. Rev. Lett. \textbf{98}, 100504 (2007).
\bibitem{Yang} W. Yang and R. B. Liu, Universality of Uhrig dynamical decoupling for suppressing qubit pure dephasing and relaxation, Phys. Rev. Lett. \textbf{101}, 180403 (2008).
\bibitem{West2010} J. R. West, B. H. Fong, and D. A. Lidar, Near-Optimal dynamical decoupling of a qubit, Phys. Rev. Lett. \textbf{104}, 130501 (2010).
\bibitem{Wang2011} Z. Y. Wang, and R. B. Liu, Protection of quantum systems by nested dynamical decoupling, Phys. Rev. A \textbf{83}, 022306 (2011).
\bibitem{Gong} M. Mukhtar, T. B. Saw, W. T. Soh, and J. Gong, Universal dynamical decoupling: Two-qubit states and beyond, Phys. Rev. A \textbf{81}, 012331 (2010).
\bibitem{Gong2010} M. Mukhtar, W. T. Soh, T. B. Saw, and J. Gong, Protecting unknown two-qubit entangled states by nesting Uhrig's dynamical decoupling sequences, Phys. Rev. A \textbf{82}, 052338 (2010).
\bibitem{Liu2010} K. Chen and R. B. Liu, Dynamical decoupling for a qubit in telegraphlike noises, Phys. Rev. A \textbf{82}, 052324 (2010).
\bibitem{Liu2011} Z. Y. Wang and R. B. Liu, Extending quantum control of time-independent systems to time-dependent systems, Phys. Rev. A \textbf{83}, 062313 (2011).
\bibitem{Biercuk} M. J. Biercuk, H. Uys, A. P. VanDevender, N. Shiga, W. M. Itano, and J. J. Bollinger, Optimized dynamical decoupling in a model quantum memory, Nature \textbf{458}, 996 (2009).
\bibitem{Du} J. Du, X. Rong, N. Zhao, Y. Wang, J. Yang, and R. B. Liu, Preserving electron spin coherence in solids by optimal dynamical decoupling, Nature \textbf{461}, 1265 (2009).
\bibitem{Ryan} C. A. Ryan, J. S. Hodges, and D. G. Cory, Robust decoupling techniques to extend quantum coherence in diamond, Phys. Rev. Lett. \textbf{105}, 200402 (2010).
\bibitem{deLange} G. deLange, Z. H. Wang, D. Rist\'{e}, V. V. Dobrovitski, and R. Hanson, Universal dynamical decoupling of a single solid-state spin from a spin bath, Science \textbf{330}, 60 (2010).
\bibitem{Wang} Y. Wang, X. Rong, P. Feng, W. Xu, B. Chong, J. H. Su, J. Gong, and J. Du. Preservation of bipartite pseudoentanglement in solids using dynamical decoupling, Phys. Rev. Lett. \textbf{106}, 040501 (2011).
\bibitem{Alvarez} G. A. \'{A}lvarez, A. M. Souza, and D. Suter, Iterative rotation scheme for robust dynamical decoupling, Phys. Rev. A \textbf{85}, 052324 (2012).
\bibitem{Tripathi} V. Tripathi, H. Chen, M. Khezri, K. W. Yip, E. M. L. Falk, and D. A. Lidar, Suppression of crosstalk in superconducting qubits using dynamical decoupling, Phys. Rev. Appl. \textbf{18}, 024068 (2022).
\bibitem{Ezzell} N. Ezzell, B. Pokharel, L. Tewala, G. Quiroz, and D. A. Lidar, Dynamical decoupling for superconducting qubits: A performance survey, Phys. Rev. Appl. \textbf{20}, 064027 (2023).
\bibitem{Zanardi} P. Zanardi, Symmetrizing evolution, Phys. Lett. A \textbf{258}, 77 (1999).
\bibitem{Viola2000} L. Viola, E. Knill, and S. Lloyd, Dynamical generation of noiseless quantum subsystems, Phys. Rev. Lett. \textbf{85}, 3520 (2000).
\bibitem{Lidar2008} D. A. Lidar, Towards fault tolerant adiabatic quantum computation, Phys. Rev. Lett. \textbf{100}, 160506 (2008).
\bibitem{West} J. R. West, D. A. Lidar, B. H. Fong, and M. F. Gyure, High fidelity quantum gates via dynamical decoupling, Phys. Rev. Lett. \textbf{105}, 230503 (2010).
\bibitem{Byrd} M. S. Byrd and D. A. Lidar, Comprehensive encoding and decoupling solution to problems of decoherence and design in solid-state quantum computing, Phys. Rev. Lett. \textbf{89}, 047901 (2002).
\bibitem{Ng} H. K. Ng, D. A. Lidar, and J. Preskill, Combining dynamical decoupling with fault-tolerant quantum computation, Phys. Rev. A \textbf{84}, 012305 (2011).
\bibitem{Zhao} P. Z. Zhao, X. Wu, and D. M. Tong, Dynamical-decoupling-protected nonadiabatic holonomic quantum computation, Phys. Rev. A \textbf{103}, 012205 (2021).
\bibitem{Xu} G. F. Xu, and G. L. Long, Protecting geometric gates by dynamical decoupling, Phys. Rev. A \textbf{90}, 022323 (2014).
\bibitem{Wu2020} X. Wu and P. Z. Zhao, Universal nonadiabatic geometric gates protected by dynamical decoupling, Phys. Rev. A \textbf{102}, 032627 (2020).
\bibitem{Wu} C. F. Wu, C. F. Sun, J. L. Chen, and X. X. Yi, Decoherence-Protected implementation of quantum gates, Phys. Rev. Appl. \textbf{19}, 034069 (2023).
\bibitem{Souza} A. M. Souza, G. A. \'{A}lvarez, and D. Suter, Experimental protection of quantum gates against decoherence and control errors, Phys. Rev. A \textbf{86}, 050301(R) (2012).
\bibitem{Suter} D. Suter and G. A. \'{A}lvarez, Protecting quantum information against environmental noise, Rev. Mod. Phys. \textbf{88}, 041001 (2016).
\bibitem{Zhang} J. F. Zhang, A. M. Souza, F. D. Brandao, and D. Suter, Protected quantum computing: Interleaving gate operations with dynamical decoupling sequences, Phys. Rev. Lett. \textbf{112}, 050502 (2014).
\bibitem{Viola2009} K. Khodjasteh and L. Viola, Dynamically error-corrected gates for universal quantum computation,Phys. Rev. Lett. \textbf{102}, 080501 (2009).
\bibitem{ViolaPRA} K. Khodjasteh and L. Viola, Dynamical quantum error correction of unitary operations with bounded controls, Phys. Rev. A \textbf{80}, 032314 (2009).
\bibitem{Viola2010} K. Khodjasteh, D. A. Lidar, and L. Viola, Arbitrarily accurate dynamical control in open quantum systems, Phys. Rev. Lett. \textbf{104}, 090501 (2010).

\bibitem{Hart} T. P. Harty, D. T. C. Allcock, C. J. Ballance, L. Guidoni, H. A. Janacek, N. M. Linke, D. N. Stacey, and D. M. Lucas, High-fidelity preparation, gates, memory, and readout of a trapped-ion quantum bit, Phys. Rev. Lett. \textbf{113}, 220501 (2014).
\bibitem{Barends} R. Barends, J. Kelly, A. Megrant, A. Veitia, D. Sank, E. Jeffrey, T. C. White, J. Mutus, A. G. Fowler, B. Campbell, Y. Chen, Z. Chen, B. Chiaro, A. Dunsworth, C. Neill, P. O' Malley, P. Roushan, A. Vainsencher, J. Wenner, A. N. Korotkov, A. N. Cleland and J. M. Martinis, Superconducting quantum circuits at the surface code threshold for fault tolerance, Nature \textbf{508}, 500 (2014).
\bibitem{Evered} S. J. Evered, D. Bluvstein, M. Kalinowski, S. Ebadi, T. Manovitz, H. Zhou, S. H. Li, A. A. Geim, T. T. Wang, N. Maskara, H. Levine, G. Semeghini, M. Greiner, V. Vuleti\'{c}, and M. D. Lukin, High-fidelity parallel entangling gates on a neutral-atom quantum computer. Nature \textbf{622}, 268 (2023).
\bibitem{Zhang2015} J. F. Zhang and D. Suter, Experimental protection of two-qubit quantum gates against environmental noise by dynamical decoupling, Phys. Rev. Lett. \textbf{115}, 110502 (2015).
\bibitem{Sar} T. van der Sar, Z. H. Wang, M. S. Blok, H. Bernien, T. H. Taminiau, D. M. Toyli, D. A. Lidar, D. D. Awschalom, R. Hanson, and V. V. Dobrovitski, Decoherence-protected quantum gates for a hybrid solid-state spin register, Nature \textbf{484}, 82 (2012).
\bibitem{LiuNC2013} G. Liu, H. C. Po, J. Du, R. B. Liu, and X. Y. Pan, Noise-resilient quantum evolution steered by dynamical decoupling, Nat. Commun. \textbf{4}, 2254 (2013).
\bibitem{IBMQuantum} IBM Quantum (2024), https://quantum.ibm.com/.
\bibitem{Krantz} P. Krantz, M. Kjaergaard, F. Yan, T. P. Orlando, S. Gustavsson, and W. D. Oliver, A quantum engineer's guide to superconducting qubits, Appl. Phys. Rev. \textbf{6}, 021318 (2019).
\bibitem{Mousolou} V. A. Mousolou, C. M. Canali, and E. Sj\"{o}qvist, Universal non-adiabatic holonomic gates in quantum dots and single-molecule magnets, New J. Phys. \textbf{16}, 013029 (2014).
\bibitem{Doherty} M. W. Doherty, N. B. Manson, P. Delaney, F. Jelezko, J\"{o}rg Wrachtrup, and L. C.L. Hollenberg, The nitrogen-vacancy colour centre in diamond, Phys. Rep. \textbf{528}, 1 (2013).
\bibitem{Abhari} A. J. Abhari, M. Treinish, K. Krsulich, C. J. Wood, J. Lishman, J. Gacon, S. Martiel, P. D. Nation, L. S. Bishop, A. W. Cross, B. R. Johnson, and J. M. Gambetta, Quantum computing with Qiskit, arXiv:2405.08810.
\bibitem{Zeng} J. Zeng, C. H. Yang, A. S. Dzurak, and E. Barnes, Geometric formalism for constructing arbitrary single-qubit dynamically corrected gates, Phys. Rev. A \textbf{99}, 052321 (2019).
\bibitem{Dong} W. Dong, F. Zhuang, S. E. Economou, and E. Barnes, Doubly geometric quantum control, PRX Quantum \textbf{2}, 030333 (2021).
\bibitem{Yi} K. Yi, Y. J. Hai, K. Luo, J. Chu, L. Zhang, Y. Zhou, Y. Song, S. Liu, T. Yan, X. H. Deng, Y. Chen, and D. Yu, Robust quantum gates against correlated noise in integrated quantum chips, Phys. Rev. Lett. \textbf{132}, 250604 (2023).
\bibitem{Montangero} S. Montangero, T. Calarco, and R. Fazio, Robust optimal quantum gates for Josephson charge qubits, Phys. Rev. Lett. \textbf{99}, 170501 (2007).
\bibitem{Rebentrost} P. Rebentrost, I. Serban, T. Schulte-Herbr\"{u}ggen, and F. K. Wilhelm, Optimal control of a qubit coupled to a mon-Markovian environment, Phys. Rev. Lett. \textbf{102}, 090401 (2009).
\bibitem{Tsai} D. B. Tsai, P. W. Chen, and H. S. Goan, Optimal control of the silicon-based donor-electron-spin quantum computing, Phys. Rev. A \textbf{79}, 060306(R) (2009).
\bibitem{Motzoi} F. Motzoi, M. Gambetta, S. T. Merkel, and F. K. Wilhelm, Optimal control methods for rapidly time-varying Hamiltonians, Phys. Rev. A \textbf{84}, 022307 (2011).
\bibitem{Huang} C. H. Huang, and H. S. Goan, Robust quantum gates for stochastic time-varying noise, Phys. Rev. A \textbf{95}, 062325 (2017).
\bibitem{Larrouy} A. Larrouy, S. Patsch, R. Richaud, J. M. Raimond, M. Brune, C. P. Koch, and S. Gleyzes, Fast navigation in a large Hilbert space using quantum optimal control, Phys. Rev. X \textbf{10}, 021058 (2020).
\bibitem{Calzavara} M. Calzavara, Y. Kuriatnikov, A. Deutschmann-Olek, F. Motzoi, S. Erne, A. Kugi, T. Calarco, J. Schmiedmayer, and M. Pr\"{u}fer, Optimizing optical potentials with physics-inspired learning algorithms, Phys. Rev. Appl. \textbf{19}, 044090 (2023).
\bibitem{Hailski} A. Halaski, M. G. Krauss, D. Basilewitsch, and C. P. Koch, Quantum optimal control of squeezing in cavity optomechanics, Phys. Rev. A \textbf{110}, 013512 (2024).
\bibitem{Motzoi2} B. Li, T. Calarco, and F. Motzoi, Experimental error suppression in Cross-Resonance gates via multi-derivative pulse shaping, npj Quantum Information \textbf{10}, 66 (2024).
\bibitem{Zahedinejad} E. Zahedinejad, J. Ghosh, and B. C. Sanders, Designing high-fidelity single-shot three-qubit gates: A machine-learning approach, Phys. Rev. Appl. \textbf{6}, 054005 (2016).
\bibitem{Niu} M. Y. Niu, S. Boixo, V. N. Smelyanskiy, and H. Neven, Universal quantum control through deep reinforcement learning, npj Quantum Inf. \textbf{5}, 33 (2019).
\bibitem{Fillingham} Z. Fillingham, H. Nevisi, and S. Dora, Optimisation of pulse waveforms for qubit gates using deep learning, arXiv:2408.02376.
\bibitem{Liu2013} Z. Y. Wang, and R. B. Liu, No-go theorems and optimization of dynamical decoupling against noise with soft cutoff, Phys. Rev. A \textbf{87}, 042319 (2013).: IBM Quantum.
\bibitem{Gong11} Y. Pan, Z. R. Xi, and J. Gong, Optimized dynamical decoupling sequences in protecting two-qubit states, J. Phys. B \textbf{44},  175501 (2011).

\bibitem{Jurcevic2021} P. Jurcevic et al., Demonstration of quantum volume 64 on a superconducting quantum computing system, Quantum Sci. Technol. {\bf 6}, 025020 (2021).
\bibitem{Gambetta2020} N. Sundaresan, I. Lauer, E. Pritchett, E. Magesan, P. Jurcevic, and J. M. Gambetta, Reducing Unitary and Spectator Errors in Cross Resonance with Optimized Rotary Echoes, PRX Quantum {\bf 1}, 020318 (2020).


\end{thebibliography}
\end{document}